\documentclass[12pt, draftclsnofoot, onecolumn]{IEEEtran}
\usepackage[font=small,skip=0.5pt]{caption}
\usepackage{multirow}
\usepackage{textcomp}
\usepackage{enumerate}
\usepackage{amsmath}
\usepackage{amssymb}
\usepackage{algorithmic}
\usepackage{amsfonts}
\usepackage{amsbsy}
\usepackage{graphicx}

\usepackage{epstopdf}
\usepackage{mathtools}
\usepackage{etoolbox}
\usepackage{soul}
\makeatletter
\usepackage[noadjust]{cite}
\usepackage{xcolor,cite,etoolbox}
\pretocmd\@bibitem{\color{black}\csname keycolor#1\endcsname}{}{\fail}
\newcommand\citecolor[1]{\@namedef{keycolor#1}{\color{red}}}
\makeatother
\makeatletter
\patchcmd{\@maketitle}
{\addvspace{0.5\baselineskip}\egroup}
{\addvspace{-1\baselineskip}\egroup}
{}
{}
\begin{document}

\title{UAV-assisted C-RAN for On-demand Cellular Coverage: Opportunities and Challenges }

\author{\IEEEauthorblockN{Byomakesh Mahapatra, Deepika Gupta, and Pankaj Kumar Sharma} 
	\thanks{\textit{Byomakesh Mahapatra and Pankaj Kumar Sharma are with the National Institute of Technology Rourkela, India.}}
	\thanks{\textit{Deepika Gupta is with the Dr. Shyama Prasad Mukherjee International Institute of Information Technology Naya Raipur, India.}}

		}

\maketitle

\begin{abstract}
The deployment of beyond fifth-generation (5G) infrastructure over disaster-affected regions, temporary hotspot situations (e.g., massive gatherings, etc.), complex terrains (e.g., sea, hills, marshes, etc.) poses numerous challenges for cellular service providers. Recently, unmanned aerial vehicles (UAVs) have emerged as potential candidates to overcome the aforementioned technical issues based on their multi-role capabilities to serve as aerial base stations, mobile relays, and flying wireless access points. As such, the UAVs can act as portable platforms that can be deployed immediately on demand without requiring massive ground infrastructure to support wireless services. This article introduces the integration of UAVs to cloud radio access networks (C-RAN) for beyond 5G applications. The article mainly focuses on the underlying opportunities and challenges to realize the UAV-assisted C-RAN (UC-RAN) architecture in view of three generic application scenarios, i.e., disaster management, hotspots, and complex terrains. A preliminary performance analysis via simulation is further provided for the proposed UC-RAN under hotspot application scenario based on the relevant metrics.  

\end{abstract}



\section{Introduction}
\IEEEPARstart{C}{loud} radio access network (C-RAN) architecture is of tremendous importance in fifth-generation (5G) and beyond cellular communications, especially due to its low energy consumption and implementation cost \cite{checko2014cloud_1C}.
The C-RAN is a technology that reduces computational complexity and supports multi-radio access technology (M-RAT) which is of great relevance for next-generation heterogeneous networks. However, at certain places such as complex terrains (e.g., hills, marshes, lakes, sea, etc.), deep forests, disaster regions, etc., the C-RAN encounters severe deployment challenges due to limited accessibility, unavailability of power supply, non-line-of-sight (non-LoS) problem, deep fading, etc.  To overcome these problems the cellular operators either need to deploy a larger number or a bigger size of the base stations, which directly leads to be a costly affair for them. Further, in hotspot cells where traffic at a base station (BS) exceeds a predetermined threshold value, the user equipment (UEs) may not be serviced without compromising their quality of service (QoS). Hence, there is a need of a flexible and scalable RAN platform \cite{saad2019vision} which enables on-demand coverage to ensure reliable communications in aforementioned situations. Unmanned aerial Vehicles (UAVs) have recently emerged as potential candidates for on-demand services in next-generation cellular networks due to their easy deployment as aerial BS and wireless access points \cite{Matoussi_2020}. For instance, in \cite{Kirtan2019design}, the UAV has been deployed as wireless access points using Wi-Fi under disaster situations. The UAVs can be deployed in the role of relays to extend coverage over complex terrains. Further, the UAVs can be deployed as small BSs in a hotspot cell to offload traffic from an overloaded ground BS (GBS). However, the deployment strategy to be employed for UAVs depends upon the specific application scenario. The main challenge associated with deploying UAVs is the limited battery power onboard. In recent years, i.e., from 2016 to 2022, a market growth of nearly 300\% has been observed in the overall use of UAVs related to military, commercial, and consumer applications \cite{Web1_Francesco}, \cite{Web2_Fortune}. The much-anticipated increase in the use of UAVs in future networks would require more sophisticated connection and traffic management strategies. Additionally, for delay-sensitive applications, a high data rate supporting robust communication links would be needed. 

\begin{figure}[!t]
	\centering
	\fbox{\includegraphics[scale=0.5]{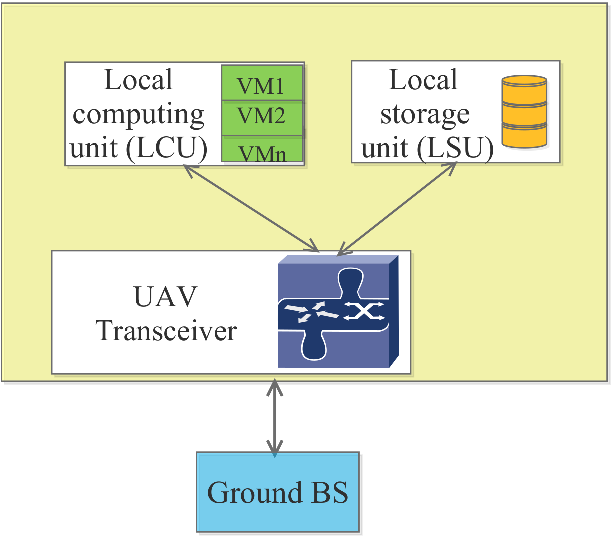}}
	\vspace{5pt}
	\caption{Mobile edge computing-enabled UAV RRH.}
	\label{fig12}
\end{figure}

Furthermore, a GBS in existing cellular networks, due to is large size, incurs very high total costs in terms of capital expenditure (CAPEX) and operational expenditure (OPEX). In practice, a GBS is built upon traditional C-RAN architecture which is unscalable in nature. As such, a GBS consumes large static power (i.e., the minimum power required to operate the base station under zero traffic load condition) as compared to a small aerial BS. Therefore, this paper explores a flexible and scalable UAV-assisted C-RAN (UC-RAN) architecture for next-generation cellular networks. Note that in C-RAN, a GBS consists of two major units, namely remote radio head (RRH) and baseband Unit (BBU). Here, in UC-RAN, the UAVs acting as flying RRHs (F-RRHs) are integrated with traditional C-RAN for on-demand coverage. The specific role of F-RRH (i.e., aerial BS, relay, wireless access point, etc.) depends upon the particular application scenario. In general, an F-RRH is equipped with a transceiver unit connected to a static RRH (S-RRH) through the wireless fronthaul. This fronthaul works based on the 3rd Generation Partnership Project (3GPP) 7.2 C-RAN functional split \cite{jiang2022wireless}. The advantage of this split is that it minimizes the required fronthaul bandwidth by performing some RF-related baseband function at the F-RRH itself. Further, S-RRHs are connected BBU pool via fixed fronthaul optical links and to the UEs through RF channels. As such, the fixed S-RRH and BBU pool placement in C-RAN is not suitable for quick service recovery under on-demand coverage applications. Furthermore, for applications like disaster management, remote area surveillance, or smart city traffic monitoring there is a need for a delay-aware data delivery system to take quick action and responses. These delay-sensitive data deliveries can be obtained either by providing a larger bandwidth to the system or by bringing the computing nodes nearer to the sender node (i.e., UE). The F-RRH-enabled Mobile Edge Computing (MEC) brings computing resources to the F-RRH that process and execute the incoming task over a local edge processor. An MEC-enabled F-RRH is shown in Fig. \ref{fig12}. It consists of three units, i.e., a computing unit, a storage unit, and a transceiver unit. The computing unit is equipped with a number of virtual machines (VMs) to process the tasks assigned by the associated UEs (or other UAVs). Each VM is separated by the others through a guest operating system and can handle different tasks parallelly over a single platform. The storage comprises of low-end memory devices to store any intermediate data generated by the local computations. The computing and storage units together help in reducing the computational load on the core network and associated delay through distributed edge processing. Here, the UAV transceiver coordinates the connection between the S-RRH and UAV edge node. Backed by features such as cost-effectiveness, low computational complexity, and less energy consumption, the UC-RAN architecture incorporating the F-RRHs is of great importance for on-demand coverage.   

Motivated by the above, this article introduces the integration of UAVs to C-RAN for relevant applications in $5$G and beyond systems. Specifically, a detailed discussion is provided on the opportunities and underlying implementation challenges related to UC-RAN architecture in the context of the following generic application scenarios: disaster management, hotspots, and complex terrain. Further, a preliminary simulation-based performance analysis of the proposed UC-RAN architecture in terms of service delay, UE blocking probability, and total power consumption will be provided to draw useful insights.              

This article has been organized as follows: Section \ref{II} provides the details of UAV-assisted RAN architecture, application scenarios, and state-of-the-art. Section \ref{III} addresses the key implementation issues for UC-RAN. Section \ref{IV} discusses monitoring and control methods for considered UC-RAN applications. Section \ref{V} deals with the performance evaluation of the proposed UC-RAN architecture. Finally, a conclusion is drawn in Section \ref{VI}.      

\section{UC-RAN Architecture: Overview and State-of-the-art}\label{II}
The conventional C-RAN architecture consists of S-RRHs connected to a BBU pool via fixed point-to-point fronthaul links. The restricted fronthaul link and time-consuming S-RRH deployment limit its cellular services in some on-demand situations, e.g., natural disasters, hotspots, complex terrain, etc. So future cellular networks need a UAV BS or F-RRH as part of it to extend coverage and provide on-demand dynamic capacity of the ground base station infrastructure. The UC-RAN  architecture UAVs are mounted with a transceiver unit and it is called F-RRH. The F-RRHs possess wireless connectivity with the S-RRHs through a wireless frontahul link. Due to its deployment flexibility, an F-RRH can take on various technical roles, e.g.,  a relay node between the S-RRH and UE, a wireless access point to UEs backhauled by the S-RRH, and a mobile edge node in on-demand situations. Further, the F-RRHs in UC-RAN can be of two types, namely passive and active. In particular, a passive F-RRH has limited functionality as it performs only simple RF signal processing such as uplink/downlink conversion and data framing. Therefore, the passive F-RRHs are primarily suited for deployment either as a relay node or a wireless access point in UC-RAN. On the contrary, an active F-RRH is enabled by the mobile edge cloud (MEC) computing platform to perform both the RF and baseband signal processing onboard the edge node. Hence, the MEC-enabled F-RRH can act as an autonomous aerial BS which can reduce the associated delay and dependency on BBU pool. Moreover, by incorporating control and management functionalities at the F-RRH edge node, it becomes a suitable candidate for many mission-critical and disaster management application scenarios. 
Table \ref{tab} highlights some key benefits and limitations of UC-RAN architecture keeping in view the traditional C-RAN architecture. Here, we mainly included the parameters that are closely related to the on-demand deployment of base stations for rapid cellular connectivity over the target areas. Table \ref{tab} reveals that the UC-RAN architecture is advantageous in terms of enhanced coverage, capacity, deployment flexibility, etc. However, the penalty incurred due to the use of F-RRHs is the increased complexity and reduced robustness of UC-RAN architecture for on-demand requirements.

\begin{figure}[!t]
	\centering
	\fbox{\includegraphics[scale=0.4]{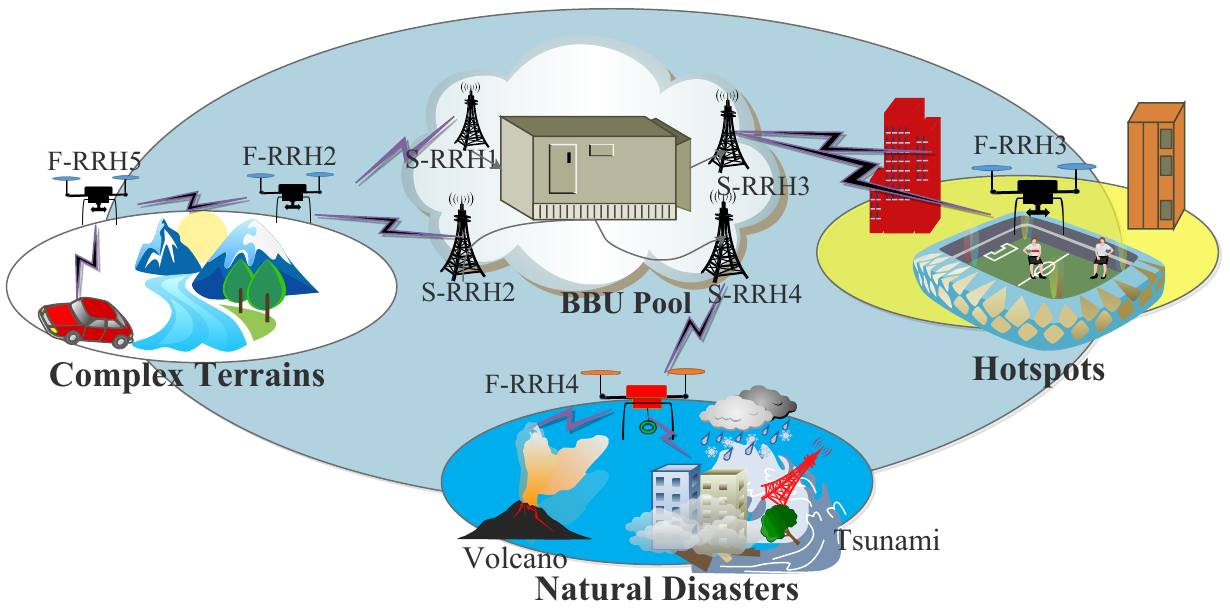}}
	\vspace{5pt}
	\caption{UC-RAN architecture for typical application scenarios.}
	\label{fig1}
\end{figure}

\begin{table*}[!t]
	\renewcommand{\arraystretch}{1.5}
	\caption{Key advantages and limitations of UC-RAN  architecture}
	\label{tab1}
	\centering
	\begin{tabular}{c|c}
		\hline\hline
	 \textbf{Advantage(s)}&  \textbf{Limitation(s)} \\
  \hline\hline
		Support on-demand RRH deployment  & F-RRH is not robust        \\
		\hline
		Support on-demand coverage enhancement  & F-RRH operational time is limited         \\
		\hline
		Support multi-altitude F-RRH placement & Increases BS complexity   \\
		\hline
		Required less CAPEX  for on-demand RRH placement  & F-RRH durability is less \\
		\hline
		
	\end{tabular}\label{tab} 
\end{table*} 
A UC-RAN architecture is shown in Fig.~\ref{fig1} for typical on-demand applications. These applications include scenarios such as cellular coverage extension in complex terrains, user management in hotspots, disaster management, and fire-fighting operations. 
Here, the UC-RAN architecture uses either passive or active F-RRHs based on the application scenario. For instance, in complex terrains, the F-RRH acts as an intermediary relay node between the S-RRH and UEs. Similarly, in a hotspot situation, there is a need of a temporary wireless access point to serve UEs facing network congestion from overloaded BS. This can be resolved by temporarily deploying some passive F-RRHs within the coverage of an S-RRH. These temporary F-RRHs effectively shed the load from the congested BS by serving the UEs. Furthermore, in disaster situations when the ground S-RRH gets damaged, both the passive and active F-RRH deployment is necessary for quick service recovery. Here, the active F-RRHs act as central nodes that control various associated passive F-RRHs engaged in disaster recovery and relief operations. A few works are available in the literature that have considered UAV-assisted BS for on-demand situations. The authors in \cite{kishk2020_teatherRAN} have proposed a tethered-UAV architecture. The tether link between UAV and GBS helps to increase flight time due to the availability of continuous power supply but limits the UAV coverage area. 
In \cite{li2017_hetrogeneouRAN}, the authors have considered UAV-assisted BS for dynamic coverage in a heterogeneous cellular network. The article discusses various UAV deployment strategies by considering interference and
frequency reuse.

The state-of-the-art for UC-RAN for B$5$G is currently in nascent stage. The authors in \cite{li2020data_ref1} have proposed a UC-RAN architecture where an RRH unit is mounted onto a UAV, so that it can work as a BS. The authors have developed optimal strategies for user association, power allocation, UAV locations for data rate maximization in a cellular cell. Further, the authors in \cite{ahmad2019} have presented a feasibility study for integrating the UAVs to C-RAN architecture. Here, the UAV is considered as a substitute BS to serve under the breakdown of the main BS. To enhance the coverage, the rate splitting technique was considered between UAV and BS. In \cite{horani2018_ref3}, the authors have compared the service latency for traffic assignment in an MEC-enabled UAV network and a macro BS-based cellular network. It was shown that UAV-based network outperforms the macro BS-based cellular network in terms of service latency. Additionally, it was depicted that integrating UAVs to existing cellular infrastructure makes the BS scalable and energy-efficient. A hybrid UC-RAN architecture with MEC is proposed in \cite{zhou2020_ref5}. Here, the authors aim to minimize the transmit power based on optimizing the user association, computational resources, and UAV location. The authors in \cite{yang2020_ref6} have proposed a UAV-assisted multi-tier MEC architecture for internet-of-things (IoT) applications. Here, the multiple UAVs have been used to serve ground IoT devices. The authors have applied a differential algorithm for optimal UAV placement and the deep reinforcement learning method for task assignment. In \cite{yang2019_ref4}, the authors have applied compressive sensing and convex optimization techniques for user association, computational resource allocation, and location planning in a UAV-assisted MEC network. In \cite{zhang2020_ref7},  an MEC-enabled UAV network was proposed for disaster management and rescue operations. It consists of two layers of UAVs with different processing and controlling capabilities. Further,  the authors have optimized the utilization of computational and communication resources in multi-tier UAV networks.   


\section{ Key Implementation Issues for UC-RAN}\label{III}
This section addresses the various key implementation issues for UC-RAN.

\subsection{Deployment Issues}
The deployment of UAVs in UC-RAN poses several challenges which are described below:
    
\textbf {Dependency on target application scenarios:} The deployment of UAVs in UC-RAN architecture depends on the target application scenarios. For instance, the deployment in disaster situations targets the search and rescue operations, supply of essential commodities, medical aid, etc. Here, an event-based UAV communications infrastructure needs to be employed. This can be technically achieved by the use of multi-hop UAV communications where an intermediate centralized F-RRH provides connectivity to other F-RRHs engaged in disaster management. It is worth mentioning that the centralized F-RRH is of active type with wireless backhaul connectivity to the S-RRH of a nearby undamaged BBU pool. Further, for certain applications such as agricultural land survey, wildlife monitoring, and reconnaissance missions, a schedule-based UAV communications infrastructure is essential. Here, the F-RRHs require occasional connectivity which can be supported by the available cellular network infrastructure. However, for deployment in complex terrains, smart cities, industrial and healthcare monitoring, etc., a permanent UAV communications infrastructure is important which poses new technical challenges.
       
\textbf{Dependency on weather conditions:} The deployment of UAVs in UC-RAN further depends on the weather conditions at a particular geographical region. The weather conditions result in various hazards to UC-RAN, viz., moderate, adverse and severe \cite{allouch2019hazard}. The moderate hazards involve poor visibility due to heavy clouds, fog, glare from sun, etc. for the physical deployment of F-RRHs. Furthermore, the adverse hazards occur due to heavy wind, turbulence, heavy rain, and snow fall at the target geographical region. This type of hazards causes problems such as the loss of communications due to unstable aerodynamic control of UAVs (e.g., drifting, wobbling, crashing, etc.). More importantly, the severe hazards are most disastrous among all which occur due to the lighting, hail, tornadoes, etc. The severe hazards are most disastrous as they can cause damage not only to the F-RRHs but also to their controlling unit and ground BS. 
 
\textbf {Dependency on altitude and LoS:} The altitude of F-RRHs is a key parameter for their safe and efficient operation. The altitude of UAVs in general is limited by the government regulations (e.g., for safer operation of commercial flight, etc.) as well as the terrain characteristics of target geographical region. Further, the LoS is the maximum range upto which the F-RRHs can see the S-RRHs at ground cellular infrastructure. The path loss in communication between F-RRHs and S-RRHs strongly depends on the LoS. With the increase in altitude of F-RRHs up to certain optimal point, the LoS improves and path loss decreases. However, beyond the optimal point, the LoS diminishes and path loss increases drastically. As mentioned previously, the altitude of F-RRHs is dependent on the terrain characteristics of a geographical region, the optimal altitude of F-RRHs is a critical factor for reliable communications and safer operation \cite{sekander2018multi}.     
\subsection{Design Issues}
The main design issues in UC-RAN are described below: 

\textbf{Payload and Weight of UAVs:} Payload is referred to as the maximum weight of equipment a UAV can carry during flight. The equipment includes the electronic circuitry, batteries, antennas, etc. The size of F-RRHs is directly related to their payload capacity. For instance, increasing the size of F-RRHs limits their endurance due to increased weight and shortened battery lifetime. Hence, the F-RRHs with discharged batteries need to be replaced frequently which require efficient admission control and handover strategies.    

\textbf{Power consumption:} The flying time of F-RRHs is limited by their battery size and the minimization of power consumption in UC-RAN is a key issue. The available power onboard the F-RRHs is utilized to power propellers for mechanical flight and to power electronic circuits for communications. The major power consumption in F-RRHs takes place in the mechanical flight which can only be reduced by a better aerodynamic design. Whereas, the reduction of electronic power consumption can be achieved by incorporating the architectural and functional changes at F-RRHs and the BBU pool in UC-RAN. Power consumption at the  BBU pool can be reduced by using centralization and virtualization at F-RRHs edge nodes. The electronic power at F-RRHs can also be reduced by an energy-efficient communication design. 

\textbf{Service latency:} It is apparent that a UC-RAN involve multi-hop communication between S-RRHs and F-RRHs for various critical applications. In general, most of these applications are delay-sensitive which requires minimal end-to-end (E2E) latency for reliable communications. The latency minimization in UC-RAN is also a critical design issue. A possible solution for E2E latency reduction in UC-RAN is to incorporate edge computing at F-RRH. The local computations at edge nodes help reducing the overall E2E latency. 

\textbf{UE blocking probability :} UE Blocking probability in UC-RAN is the chance that a UE fails to connect with an S-RRH or F-RRH within a stipulated time period. In a cellular hotspot, the limited bandwidth availability at the S-RRH restricts some of the UEs to get serviced. This causes the increase in UEs' blocking probability at the S-RRH. In UC-RAN, this problem can be effectively tackled by the F-RRHs which act as temporary access points to offload UEs from an overloaded S-RRH in a hotspot cell.

\textbf{Hovering time:} Hovering time of F-RRHs is an important metric for multi-user communication scenario. Here, an F-RRH acts as a wireless access point for serving multiple groups of users in a cellular cell. The F-RRH sequentially serves multiple group of UEs by hovering for optimal time at certain strategic locations. The hovering time optimization of F-RRHs targeting the sum-rate maximization of associated UEs is an important design problem for UC-RAN. 
\section{Monitoring and Control Strategies for F-RRHs in UC-RAN}\label{IV}
The monitoring and control strategies for F-RRHs are presented for the three generic application scenarios as described previously. 
\begin{figure*}[!t]
	\centering
	\fbox{\includegraphics[scale=0.38]{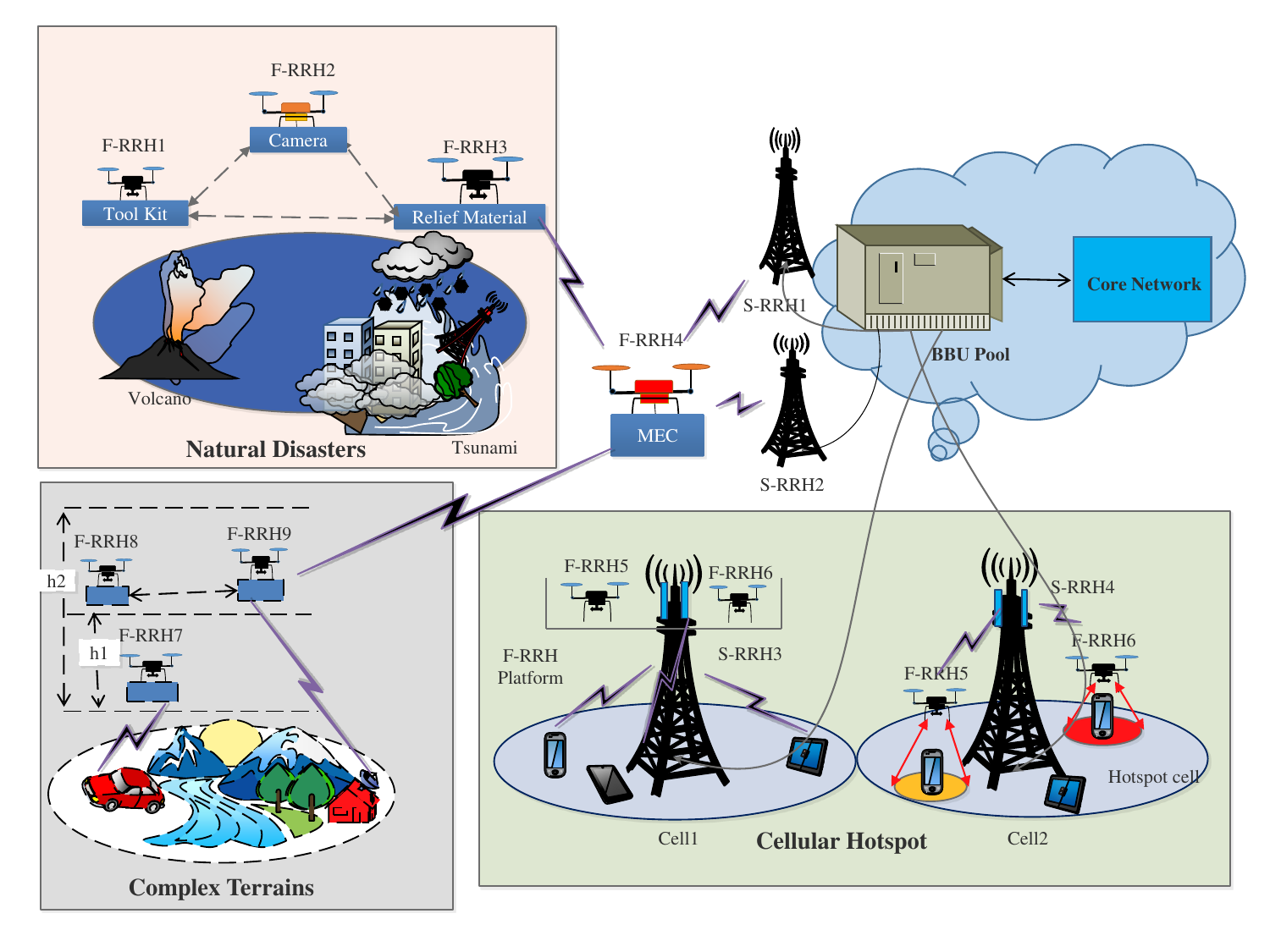}}
	\vspace{5pt}
	\caption{Implementation of UC-RAN for typical application scenarios.}
	\label{fig2}
\end{figure*}

\subsection{Disaster Management}
 In disaster situations, e.g., earthquake, tsunami, cyclones, etc., there is an immediate need to establish communications to effectively perform the relief operations. For this purpose, as shown in Fig. \ref{fig2}, a group of F-RRHs can be employed to provide wireless connectivity in the disaster hit area with a nearby working BBU pool. These F-RRHs take on various roles with dedicated wireless links as shown in the figure. For instance, F-RRH2 equipped with a camera performs the geographical survey of disaster hit region. The F-RRH3 and F-RRH4 are considered to be the intermediate relay nodes that connect the BBU pool to F-RRH1 and F-RRH2. The F-RRH4 is assumed to be an MEC-enabled node with a local control unit for connected F-RRHs which manages their inter-node link communications. As such, from a network's perspective, the resulting objective in such MEC-enabled systems is to reduce the E2E delay in communications between F-RRHs and the BBU pool. Note that the E2E delay comprises of two components, i.e., communication and processing delays. If the information from a connected F-RRH is processed by F-RRH4, the total delay can be expressed as 
\begin{equation}
    \mathcal D_{total}^{rrh} = \mathcal T_{C}^{rrh} + \mathcal D_{T}^{rrh},
\end{equation}
where $\mathcal T_{C}^{rrh}$ and $\mathcal D_{T}^{rrh}$ denote the communication and processing delays, respectively, at F-RRH1. 
Likewise, if the information received by F-RRH1 is forwarded to the BBU pool, the total delay can be expressed as
\begin{equation}
    \mathcal D_{total}^{bbu} = \mathcal T_{C}^{bbu}  + \mathcal D_{T}^{bbu},
\end{equation}
where $\mathcal T_{C}^{bbu}$ and $\mathcal D_{T}^{bbu}$ is the communication and processing delay, respectively, at the BBU.

\subsection{Hotspot}
The hotspot is a situation when the number of active users in a given cell increases beyond a threshold such that the additional users cannot be supported by the BS. The hotspots occur during the large sporting and social events in a geographical region. In hotspot situations, due to its fixed architecture, the traditional C-RAN poses several physical and economic limitations to install additional number of S-RRHs to meet increased traffic demand. Hence, a cost-effective solution to this problem is to deploy the F-RRHs. The deployment of F-RRHs in a hotspot region is controlled by the control unit housed in the BBU pool. The decision for placement of F-RRHs in a hotspot is taken by the BBU pool based on the current user density around the S-RRH and corresponding resource (bandwidth) utilization. The BBU pool regularly monitors the status of resource utilization in a cell for the connected users. As soon as it goes beyond a predetermined threshold value, an overload condition is detected at the S-RRH of the cell and the deployment of F-RRHs gets initiated by the control unit at BBU pool. Note that, hereby, the F-RRHs share the traffic load of additional users in the cell and offload the overloaded S-RRH.   

Fig. \ref{fig2} further illustrates the offloading of users by the F-RRHs in hotspot situations. Here, two different cells with individual S-RRHs controlled and monitored by a shared BBU pool are depicted. The S-RRH3 in Cell 1 has a standby platform with charging stations for F-RRHs to be deployed in hotspot. Each of these F-RRHs is assumed to be equipped with a BS transceiver units, and they are connected to the BBU pool via backhaul microwave links. The Cell 2 with S-RRH4 is considered as a hotspot. In an event when the control unit at the BBU pool detects hotspot, the F-RRHs from Cell 1 are immediately engaged at appropriate locations in Cell 2 to form small cells as shown in the figure. The users lying in these small cells are offloaded from S-RRH4 and are served by the corresponding F-RRHs. When the load at S-RRH4 reduces the engaged F-RRHs are called back and placed again in the standby mode at S-RRH3. The placement of F-RRHs in a given cell depends upon a utilization factor (UF) at an S-RRH, which is calculated based on its current load and capacity as
\begin{equation}
\ UF_{rrh}  = \frac{L_{rrh}}{C_{rrh}}\times 100,
\end{equation}
where $L_{rrh}$ denotes the traffic load at the S-RRH generated due to additional UEs and $C_{rrh}$ denotes the capacity in terms of bandwidth available at the S-RRH. 

\subsection{Complex Terrains} 
In traditional C-RAN architecture, the cellular coverage over a large complex terrain requires the installation of large number of S-RRHs interconnected via a backhaul network. However, the deployment of the large number of S-RRHs is costly over the complex terrains. In extreme terrains, i.e., marshes, lakes, etc., the ground-based infrastructure may be even infeasible to install. Further, in some terrains, e.g., hills, etc., the uneven elevations of landscape cause LoS problem for reliable communications. Fig. \ref{fig2} shows a cooperative UC-RAN architecture where the cellular coverage to a complex terrain is extended by the multiple inter-connected F-RRHs. Here, a number of F-RRHs is deployed in small clusters over a geographical region. For illustration purpose, a single cluster is shown in Fig. \ref{fig2}. Each such cluster has an MEC-enabled F-RRH as cluster-head which has backhaul connection with a BBU pool. The cluster-head is responsible for controlling and managing the other F-RRHs in the respective cluster. Multi-hop may be rendered by various F-RRHs in a cluster based on the cooperative relaying techniques. Furthermore, to reduce the number of fronthaul links, the cluster of F-RRH relies upon the extended-star topology. 

\section{Performance evaluation of UC-RAN architecture}\label{V}
 This section illustrates the performance of the proposed UC-RAN architecture based on simulations using the LTE-TU Vienna simulator \cite{mohsen2015simulator} and MATLAB. In particular, the hotspot application scenario as given in Fig. \ref{fig1} is considered for the performance evaluation. The simulation parameter settings are given in Table \ref{tab2}. Here, a comparative performance analysis has been carried-out between the macro BS, traditional C-RAN and UC-RAN architectures. Mainly, the service latency, blocking probability, and total power consumption are considered as performance metrics with respect to the variations in UEs' traffic at Cell 2 and the handover traffic owing to UEs' mobility from Cell 1 to Cell 2. Further, for evaluating the performance of macro BS-based architecture, a macro BS is assumed to be located in Cell 2 with a coverage area of $3$ km$^2$. The traffic load is measured in terms of the number of connected UEs to the BS. For this purpose, the number of connected UEs is varied from $10\%$ to $100\%$ of the maximum number of UEs which can be supported by the BS within available spectrum resources. The hotspot situation in Cell 2 is created while increasing the number of handover UEs. Furthermore, for the performance evaluation of traditional C-RAN architecture, an S-RRH and a BBU pool are deployed with the same coverage area as defined for macro BS. Also, the assignment of UEs to S-RRH is considered to be identical to that in the macro BS-based architecture. Moreover, for the performance evaluation of UC-RAN architecture, both the F-RRHs and S-RRHs are considered with the same traffic load as in previous architectures.
 \begin{table}[!t]
 	\renewcommand{\arraystretch}{1.3}
 	\caption{Simulation parameters.}
 	\label{tab1}
 	\centering
 	\begin{tabular}{c||c}
 		\hline\hline
 		\textbf{Parameters} & \textbf{Value } \\
 		\hline\hline
 		LTE Bandwidth   & $20$ MHz  \\
 		\hline
 		Number of UEs  & 100-1000  \\
 		\hline
 		Number of Macro-BS and S-RRHs  & $2$ \\
 		\hline
 		Number of UAVs  & $4$\\
 		\hline
 		Number of BBU pool & 1 \\
 		\hline
 		Transmit power of Macro BS and S-RRHs & $43$ dBm \\
 		\hline
 		Transmit power of F-RRHs & $30$ dBm \\
 		\hline
 		Modulation scheme & $64$-QAM \\
 		\hline
 		Height of macro BS and S-RRHs & $100$ ft\\
 		\hline
 		Altitude of F-RRHs & $100$ ft\\
 		\hline
 	\end{tabular}\label{tab2} 
 \end{table}          

\begin{figure}
	\centering
	\fbox{\includegraphics[scale=0.60]{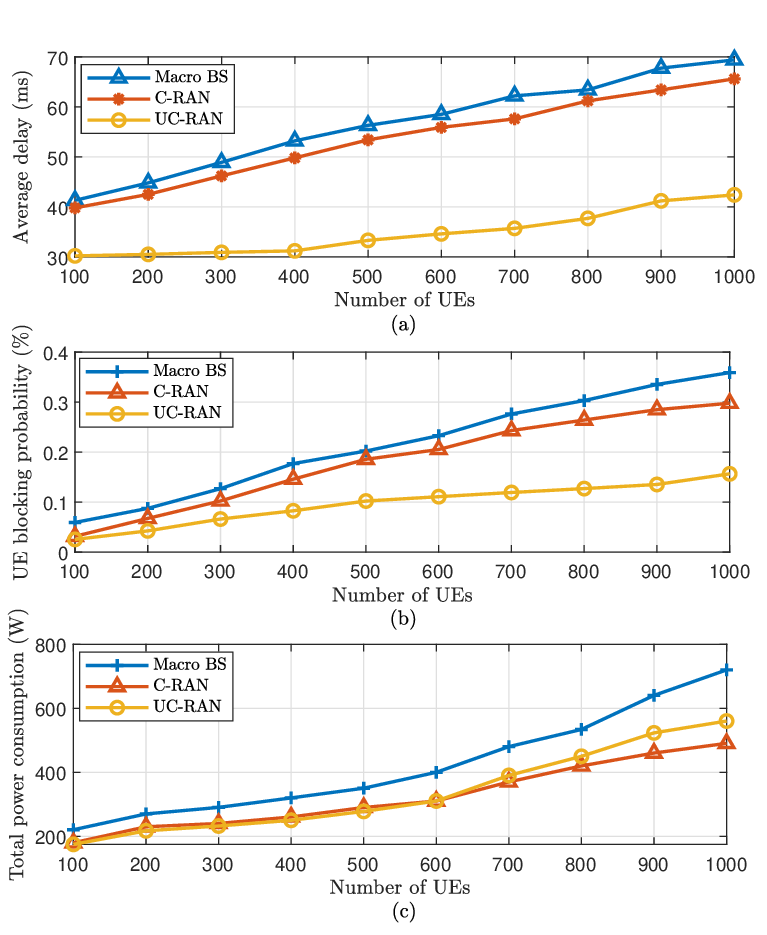}}
	\vspace{5pt}
	\caption{ Performance comparison of macro BS, C-RAN, and UC-RAN architectures in terms of: (a) average E2E delay; (b) UE blocking probability; (c) total power consumption.}
	\label{fig5}
\end{figure}

Fig. \ref{fig5}(a) plots the average E2E service delay versus number of UEs for various RAN architectures. Here, it is observed that with the increased number of UEs, the E2E average delay increases. This occurs due to the network congestion problem resulting from the increased number of UEs within the available spectrum resources at the BS in the considered RAN architectures. It is worth mentioning that among the three RAN architectures, the UC-RAN has a better and stable delay profile in hotspot application scenario even when the traffic load approaches its maximum limit at BS. The main reason behind this is the load sharing between F-RRH and S-RRH in UC-RAN as compared to an overloaded S-RRH/macro BS in other architectures.    

Fig. \ref{fig5}(b) plots the UE blocking probability versus number of UEs for the three RAN architectures. One can observe that the UE blocking probability is lowest in the UC-RAN as compared to the other RAN architectures. Here also, the improvement in UE blocking probability occurs due to the reduced load at F-RRH and S-RRH under load sharing. Nevertheless, the nearly flat nature of UE blocking probability curve for UC-RAN over a wide range of traffic load makes it useful for next-generation cellular networks.     
 
Fig. \ref{fig5}(c) plots the total power consumption versus number of UEs for the three RAN architectures. It can be observed that the total power consumption in all three architectures increases proportionally with the traffic load. Further, the total power consumption in C-RAN and UC-RAN is approximately the same for lower traffic loads. However, beyond the certain value of traffic load, the power consumption in UC-RAN relatively increases as compared to C-RAN. It is because at higher traffic loads, the power consumption at F-RRH significantly increases in addition to that at S-RRH in UC-RAN architecture. Whereas in C-RAN, the power consumption takes place only at S-RRH.       

\section{Conclusion}\label{VI}
This article has proposed a UC-RAN architecture for beyond $5$G networks. Especially, three application scenarios were considered, namely disaster management, hotspot, and complex terrain deployment. The key implementation issues along with the monitoring and control strategies of F-RRHs for UC-RAN under the aforementioned application scenarios were discussed. Further, a simulation-based comparative performance analysis was carried out for macro BS, C-RAN, and UC-RAN architectures in terms of average E2E delay, UE blocking probability, and total power consumption under the hotspot application scenario. Compared with the other two RAN architectures, the UC-RAN was found to be a promising candidate for future deployment based on its performance.   
\bibliographystyle{IEEEtran}
\bibliography{IEEEabrv,Ref11}

\end{document}